\documentclass[article,superscriptaddress,onecolumn]{revtex4}
\usepackage[ansinew]{inputenc}
\usepackage{graphicx}
\usepackage{amsmath}
\usepackage{amsthm}
\usepackage{bm}
\usepackage{layout}
\usepackage{float}
\usepackage{txfonts}
\usepackage{amsfonts}
\usepackage{amssymb}%
\usepackage{bm}
\usepackage{graphicx}

\begin{document}

\title{Information Invariance and Quantum Probabilities}\thanks{This article is dedicated to Pekka Lahti on the occasion of his 60th
birthday.}


\author{{\v C}aslav Brukner}
\affiliation{Institute of Quantum Optics and Quantum Information,
Austrian Academy of Sciences, Boltzmanngasse 3, A-1090 Vienna,
Austria} \affiliation{Faculty of Physics, University of Vienna,
Boltzmanngasse 5, A-1090 Vienna, Austria}

\author{Anton Zeilinger}
\affiliation{Institute of Quantum Optics and Quantum Information,
Austrian Academy of Sciences, Boltzmanngasse 3, A-1090 Vienna,
Austria} \affiliation{Faculty of Physics, University of Vienna,
Boltzmanngasse 5, A-1090 Vienna, Austria}

\begin{abstract}

We consider probabilistic theories in which the most elementary
system, a two-dimensional system, contains one bit of information.
The bit is assumed to be contained in any complete set of mutually
complementary measurements. The requirement of {\it invariance} of
the information under a {\it continuous} change of the set of
mutually complementary measurements uniquely singles out a measure
of information, which is {\it quadratic} in probabilities. The
assumption which gives the same scaling of the number of degrees of
freedom with the dimension as in quantum theory follows essentially
from the assumption that all physical states of a higher dimensional
system are those and only those from which one can post-select
physical states of two-dimensional systems. The requirement that no
more than one bit of information (as quantified by the quadratic
measure) is contained in all possible post-selected two-dimensional
systems is equivalent to the positivity of density operator in
quantum theory.

\end{abstract}

\maketitle

\section{Introduction: Failure of classical concepts}

\label{intro} In general quantum mechanics only makes probabilistic
predictions for individual events. Can one go beyond quantum
mechanics in this respect? With an aim to argue for incompleteness
of the quantum-mechanical description Einstein, Podolsky and
Rosen~\cite{epr} (EPR) introduced the notions of ''locality'' and
``elements of physical reality'' in their seminal paper from 1935.
In the EPR words the two notions read: (Locality) ``Since at the
time of measurements the two systems no longer interact, no real
change can take place in the second system in consequence of
anything that may be done to the first system''; and (Elements of
physical reality) ``If, without in any way disturbing a system, we
can predict with certainty (i.e., with probability equal to unity)
the value of a physical quantity, then there exists an element of
physical reality corresponding to this physical quantity.''

The theorem of Greenberger, Horne and Zeilinger~\cite{ghz} (GHZ)
showed that the mere concept of existence of local ``elements of
physical reality'' as introduced by EPR is in a contradiction with
quantum mechanical predictions. Which -- if not both -- of the two
EPR premisses is violated is an open question. One could, for
example, relax the premiss of ``elements of physical reality'' and
consider local probabilistic (also called stochastic) theories in
which the individual local result is not assumed to be
pre-determined, but only its probability to occur. To discuss what
it explicitly means consider two space-like separated parties,
colloquially called Alice and Bob, who perform measurements in their
local laboratories. In every experimental run one defines the
conditional probability (describing correlations)
$p(a,b|x,y,\lambda)$ that Alice's and Bob's outcomes are $a$ and
$b$, given that their (free~\cite{bellfree}) choices of measurement
settings are $x$ and $y$, respectively. Here $\lambda$ is the full
set of, hidden and not hidden, variables that may include all the
information about the past of both Alice and Bob except for their
choices of measurement settings. In a local probabilistic
hidden-variable theory one requires that $p(a,b|x,y,\lambda) =
p(a|x,\lambda) \cdot p(b|y,\lambda)$ is satisfied, such that
measurable conditional probability becomes $p(a,b|x,y)= \int
d\lambda \rho(\lambda) p(a|x,\lambda) \cdot p(b|y,\lambda)$, where
$\rho(\lambda)$ is the probability distribution of the hidden
variable. The theorem of Bell~\cite{bell}, which historically
precedes the GHZ theorem, shows that not even a local probabilistic
hidden-variable theory can agree with all predictions of quantum
theory. What is the lesson we can learn from this? As we here speak
about theories that treat probabilities as irreducible and which are
local, but nonetheless contradict quantum predictions, for some
authors this indicates that nature is non-local.

While the mere existence of Bohm's model~\cite{bohm} demonstrates
that non-local hidden-variables are a logically valid option, we now
know that there are plausible models, such as Leggett's
crypto-nonlocal hidden-variable model~\cite{leggett}, that are in
disagreement with both quantum predictions and
experiment~\cite{groablacher}. But, perhaps more importantly in our
view is that if one is ready to consider probabilistic theories,
then there is no immediate reason to require the locality condition
in the form given above. Violation of this condition is not in
conflict with the theory of relativity, as it does not imply the
possibility of signalling superluminally. To the contrary, quantum
correlations cannot be used to communicate from Alice to Bob, nor
from Bob to Alice, but do violate Bell's inequalities \footnote{In
relation to this it is symptomatic that quantum formalism makes no
difference between the description of two (space-like) separated
particles and of two degrees of freedom of a single particle with a
corresponding Hilbert space dimension. This suggests that spatial
distance in the ordinary three-dimensional space is irrelevant in
the abstract Hilbert space description of quantum mechanical
systems. In our view this relativizes the violation of locality
condition as a possible explanation of Bell's theorem.}. It is
therefore legitimate to consider quantum theory as a probability
theory subject to or even derivable from more general principles
such as the principle of no-signalling~\cite{popescu,barrett} or an
information-theoretical
principle~\cite{weizsaecker,wheeler,zeilinger}.

Whether or not one may be able to resolve the question which (if not
both) of the two EPR premisses is violated in quantum mechanics, one
can ask how plausible the hidden-variable program is on its own. A
particularly interesting question is what is the amount of resources
in terms of the number of hidden-variable states a hidden-variable
model must consume to be in agreement with quantum theory. In papers
of Hardy~\cite{hardyhv}, Montina~\cite{montina} and Dakic {\it et.
al}~\cite{dakic} it was shown that in the limit of large number of
measurements no hidden-variable theory that agrees with quantum
predictions could use less hidden-variable states than the
straightforward ``brute force'' model in which every quantum state
is associated with one such hidden state. This implies that no
hidden-variable theory can provide a description that is more
efficient than quantum theory itself. Even for the simplest quantum
system like electron spin or photon polarization, any
hidden-variable approach is extremely resource demanding, requiring
infinitely many hidden-variable states in order to fulfill the
requirement of explaining all possible measurements. It is
interesting here to contrast this finding with Feynman's words: ``It
always bothers me that, according to the laws as we understand them
today, it takes a computing machine an infinite number of logical
operations to figure out what goes on in no matter how tiny a region
of space and no matter how tiny a region of time, ... why should it
take an infinite amount of logic to figure out what one tiny piece
of space-time is going to do?''

Concluding the introduction we note that while maintaining the
hidden-variable program and giving up locality may logically be
possible, the analysis given above in our opinion supports Bohr's
view of~\cite{bohrepr} ``the necessity of a final renunciation of
the classical ideal of causality and a radical revision of our
attitude towards the problem of physical reality.''

\section{A foundational program based on probabilities}

Recently, several authors have been looking at ways of deriving
quantum theory from reasonable principles that are motivated by
putting primacy on the concept of information or on the concept of
probability which again can be seen as a way of quantifying
information. Particularly interesting papers which discuss this idea
are
Ref.~\cite{barrett,weizsaecker,wheeler,zeilinger,wootters81,fivel,summhammer,aage,caticha,hardy,bruknerzeilinger,fuchs,cbh,grangier,luo,grinbaum,spekkens,goyal,dariano}.
This incomplete list of various approaches should suggest that it is
possible to derive a large part of the quantum formalism without
recourse to {\it ad hoc} assumptions about the abstract Hilbert
space description of quantum systems. Rather, a goal is that the
usual rather formal mathematical axiomatization (see, for example,
Ref.~\cite{dirac}) arises as a consequence.

In this paper we consider probabilistic theories built upon the
physical principle of limited information content of any system. The
paper certainly does not present a complete axiomatic derivation of
quantum theory but rather it demonstrates that some essential
elements of the theory can be obtained from the concept of
information. We also discuss how our approach relates to others,
like those of Hardy~\cite{hardy} and Aaronson~\cite{aaronson}.

In the probabilistic theory the state of a system is defined by the
set of probabilities associated with every measurement that may be
performed on the system. Given a list of the probabilities
associated with every conceivable measurement, labeled by $i$, one
could write a state of a system as a probability vector: ${\bf
p}=(...,p_i,...)$. The list might contain superfluous information,
and instead one can look for a list that contains the minimal number
of probabilities that are sufficient to fully specify the state in
the sense that the probability of any possible result for any
conceivable measurement can be derived therefrom.

Hardy~\cite{hardy} discussed two important integers in the
reconstruction of such theories. The first one is the smallest
number $K$ of different measurements that specify the state
completely. The second one is the dimension $N$ which is the maximum
number of states that can be distinguished from one another in a
single shot experiment. Two states are distinguishable if there
exists a choice of measurement setting for which the sets of
possible outcomes they can give rise to are disjoint. Therefore, the
state is represented by a probability vector ${\bf p}=
(p_1,...,p_K)$. Wootters~\cite{wootters} and Hardy~\cite{hardy} gave
the parameter counting argument for composite systems to determine
the function $K(N)$. The argument is based on the following
assumption:

(i) {\bf (Composite Systems)} Upon combining constituents into a
composite system, dimension $N$ and number $K$ of degrees of freedom
are multiplicative, i.e. $K(N_1 N_2)=K(N_1)K(N_2)$, where $N_i$,
$i=1,2$, is dimension and $K(N_i)$ the degree of freedom of the
$i$-th constituent.

The condition leads to $K(N)=N^r-1$ with $r \in \mathbf{N}$. This
suggests the existence of a hierarchy of theories, which correspond
to different $r$~\cite{hardy,zyczkowski,paterek}.

We begin our considerations from the guiding principle
that~\cite{zeilinger}

(ii) ({\bf Limited Information}) The most elementary system
(two-dimensional system) contains one bit of information.

The most elementary system is of dimension two. The principle states
that there is a measurement with two distinguishable outcomes in
either of which the system may give a deterministic answer. When the
system gives such an answer it is called to be in a pure state. It
is clear that depending on the particular function $K(N)$ one can
have different theories in agreement with the principle. For
$K=2^1-1=1$ one has a classical bit for which only two states ``0''
and ``1'' satisfy the principle when the system definitely gives
either answer. Any convex combination (classical mixture) of the two
states, $p$``0''$+(1-p)$``1'', gives rise to a (mixed) state that
has {\it less} than one bit of information. This statement is
independent of how we quantify information, i.e. of the particular
choice of information measure $H(p,1-p)$. Whenever one has a convex
combination of two states, one loses the information from which
state a particular sample comes from.

The next most simple case is for $K=2^2-1=3$ which is the case in
quantum  theory. One needs three independent measurements to
completely specify the state of a two-dimensional system. We make
here a non-trivial assumption that one can choose mutually
complementary measurements for these three independent measurements
in a general probabilistic theory with $K=3$ (in quantum theory
these measurements are known as mutually
unbiased~\cite{woottersfields,ivanovic}). They have the property of
mutual exclusiveness: the total knowledge of one observable is at
the cost of total ignorance about the other two complementary ones.
More precisely, if probability for an outcome in one of the three
experiments is unity, then the two outcomes in either of the two
complementary experiments are equally probable. In this sense
mutually complementary measurements constitute ``maximally
independent pieces of information'' that are contained in the
system. They serve as a useful ``reference frame'' to describe how
information that is contained in a system is distributed over
independent observables. An explicit example of a set of mutually
complementary observables in quantum theory are three spin
projections along orthogonal directions of a spin-1/2 particle.
Having this example in mind we will denote three mutually
complementary measurements as $x$, $y$ and $z$. The state is thus
given by ${\bf p}=(p_x,1-p_x,p_y,1-p_y,p_z,1-p_z)$. (One can
additionally reduce the number of components in the vector since the
probabilities sum up to unity in every individual experiment. We
leave the extended form for the purposes of further discussion.).

Concluding this section we note that in a general probabilistic
theory of a two-dimensional state there are $K=2^r-1$, $r=1,2,3
...$, mutually complementary measurements~\cite{paterek}. This
number agrees with the parameter counting
arguments~\cite{wootters,hardy} mentioned above and, independently,
follows from {\it operational} considerations in which complementary
measurements reveal mutually complementary properties of black box
configurations in the framework of black-box
computation~\cite{paterek}. 

\section{Uniqueness of the quadratic measure of information}

The necessary choice between observing path information and the
observability of interference patterns is one of the most basic
manifestations of quantum complementarity, as introduced by Niels
Bohr. It does not only compromise the extreme cases of maximal
knowledge of path information at the expense of complete loss of
interference and vice versa, but also intermediate situations in
which one can obtain some partial knowledge about the particle’s
path and still observe an interference pattern of reduced contrast
as compared to the ideal interference
situation~\cite{wootterszurek,greenberger,jaeger,englert,bruknerprl}.
This and other manifestations of quantum complementarity suggests
that our total knowledge, or information, about the outcomes in
complementary experiments that can be performed on a given quantum
system is limited. This information may then manifest itself fully
as path information or as modulation of the interference pattern or
partially in both to the extent defined by the finiteness of
information. The question is how to give a formal mathematical
description to this observation.

Bohr~\cite{bohr1958} remarked that ``... phenomena under different
experimental conditions, must be termed complementary in the sense
that each is well defined and that together they exhaust all
definable knowledge about the object concerned.'' This suggests that
the total information content of a quantum system is somehow
contained in the full set of mutually complementary experiments. We
follow here a natural choice and define the total information (of 1
bit) of a two-dimensional system as the {\it sum} of the individual
measures of information over three mutually complementary
experiments. For convenience we will use here not a measure of
information or knowledge, but rather its opposite, a measure of
uncertainty or entropy. This choice has no relevance for our
findings. Therefore, the total uncertainly about a quantum system is
defined as
\begin{equation}
H_{total} = H(p_x,1-p_x) + H(p_y,1-p_y) + H(p_z,1-p_z) = 2,
\label{invariance}
\end{equation}
where $H(p_u,1-p_u)$ is some measure of uncertainty that is
associated to the probability distribution observed in experiment
$u$. At this stage of the argument it is assumed that $H$ can be an
arbitrary function of probabilities that fulfills
Eq.~(\ref{invariance}) and is normalized such that for an individual
experiment it takes its maximal value of 1 for completely random
outcomes ($p_u=1/2$) and minimal value of 0 for deterministic
results ($p_u=0$ or $1$). Note, however, that one can have minimal
uncertainty in at most one of the three mutually complementary
measurements, therefore the sum in Eq.~(\ref{invariance}) has to be
2. It should also be mentioned that in general $H_{total} \geq 2$ to
include the case of mixed states, which are defined as convex
mixtures of pure states.

In a general probabilistic theory one can analogously define the
total uncertainty as
\begin{equation}
H_{total} = \sum_{j=1}^{2^r-1} H(p_j,1-p_j) = 2^r-2, \label{perhaps}
\end{equation}
where we sum over all $2^r-1$ mutually complementary measurements
and again in at most one of them we can have full knowledge. The
quantum case~(\ref{invariance}) is recovered for $r=2$.

Which measure of uncertainty should we use to quantify uncertainties
in Eq.~(\ref{invariance}) and (\ref{perhaps})? In literature one can
find a large number of different measures. In fact, there is an
(uncountable) infinite number of generalized measures, known as
entropies of degree $\alpha$,
\begin{equation}
H_{\alpha}(p_1,...,p_n) := k \frac{1-\sum_{i=1}^{n}
p^\alpha_i}{\alpha-1}. \label{tsallis}
\end{equation}
where $k$ and $\alpha$ are two constants and $n$ is the number of
different outcomes. They were introduced by
Havrda-Charv\'{a}t~\cite{havrdacharvat} and are also known as
Tsallis~\cite{tsallis} entropies in statistical physics. They are
one-parameter generalization of the Shannon entropy \footnote{The
characteristic property of Tsallis entropies is called
pseudoadditivity: $\frac{H_\alpha(AB)}{k} = \frac{H_\alpha(A)}{k} +
\frac{H_\alpha(B)}{k} + (1-\alpha) \frac{H_\alpha(A)}{k}
\frac{H_\alpha(B)}{k}$ holds true for two mutually independent
finite event systems $A$ and $B$.},
\begin{equation}
\lim_{\alpha \rightarrow 1} H_\alpha = H_1= -k \sum_{i=1}^{n} p_i
\log p_i.
\end{equation}

We will now introduce two physically reasonable requirements on
$\alpha$-entropies, which will finally uniquely single out the
$\alpha=2$ entropy. We require (a) the total uncertainty of a system
to be independent of the particular choice of the complete set of
mutually complementary measurements, and (b) that the change from
one to another set can be performed in a continuous fashion.
Equivalent is the requirement that

(iii) ({\bf Information Invariance \& Continuity}) The total
uncertainty (or total information of one bit) is invariant under a
continuous change between different complete sets of mutually
complementary measurements.

As noted above a selection of a specific set is like a selection of
a reference frame: the total uncertainty should be independent of
the observers' choice of how he represents his knowledge about the
system. Mathematically, the property of invariance means that
Eq.~(\ref{invariance}) is valid for all sets of choices of $x$, $y$
and $z$. From the form of $\alpha$-entropies~(\ref{tsallis}) one
concludes that the invariance property is equivalent to the
condition that the {\it $\alpha$-norm of the probability vector
${\bf p}$ is preserved under the change of the set of mutually
complementary measurements.} The $\alpha$ norm of a given vector
${\bf r}= (r_1,...,r_m)$ is defined as $||{\bf r}||_\alpha :=
(\sum_{i=1}^m |r_i|^\alpha|)^{1/\alpha}$.

By choosing different set of complementary measurements, the state
${\bf p}$  will be transformed. Hence, it will go from {\bf p} to
some new state {\bf f(p)}, where {\bf f} is a vector function
associated with the transformation. Hardy~\cite{hardy} showed that
the {\bf f} is actually a linear transformation. Hence the
transformation is given by ${\bf p}\rightarrow {\hat A}{\bf p}$,
where $A$ is a real matrix \footnote{It should be noted that
non-linear transformations could lead to superluminal
signaling~\cite{gisin}, violation of the second law of
thermodynamics~\cite{peres} or solubility of NP-complete
problems~\cite{adamsloyd}, which all question their physical
justification.}.

We next use a result of Aaronson~\cite{aaronson}, which is derived
in a different context. He considered which matrices $\hat{A} \in
{\bf R}^{m \times m}$ have the property that for {\it all} vectors
${\bf r}$, $|| \hat{A} {\bf r} ||_{\alpha} = ||{\bf r} ||_{\alpha}$.
He proved that {\it if $\alpha \neq 2$, then the only
$\alpha$-norm-preserving linear transformations are permutations of
diagonal matrices.} These transformations are discrete and cannot
account for a {\it continuous} change of vector ${\bf r}$ that would
preserve its norm. If  $\alpha = 2$, the norm-preserving
transformations are orthogonal matrices that continuously change
${\bf r}$.

Taking now the probability vector ${\bf p}$ in the proof of
Ref.~\cite{aaronson} one can show that the only measure of
information that satisfies the requirement of invariance under {\it
continuous} change of the probability vector is the entropy of
degree $\alpha=2$. One should here be careful, because the
probability vectors are not arbitrarily 6-vectors as in the proof of
Ref.~\cite{aaronson} but of the form $(p_x, 1-p_x,
p_y,1-p_y,p_z,1-p_z)$ (the probabilities are nonnegative and they
sum up to unit in any of the three individual experiments). This
implies that the transformation must be {\it stochastic} in the
sectors $(p_x, 1-p_x)$, $(p_y,1-p_y)$ and $(p_z,1-p_z)$ of the
probability vector, which correspond to individual experiments $x$,
$y$ and $z$, respectively. With this additional condition fulfilled
one can show that the original statement is still valid: if $\alpha
\neq 2$, then the only $\alpha$-norm-preserving linear
transformations of probability vectors are permutations of diagonal
matrices. To guarantee that the transformation is stochastic in the
corresponding sectors of the probability vector one has that two
probabilities from one sector ($p_i,1-p_i)$ are always permuted with
two probabilities from the second sector $(p_j,1-p_j)$. An explicit
example of one such discrete transformation is as follows:
\begin{equation}
\hat{A}=
\begin{pmatrix}
0 & 0 & 1 & 0 & 0 & 0 \\
0 & 0 & 0 & 1 & 0 & 0 \\
0 & 1 & 0 & 0 & 0 & 0 \\
1 & 0 & 0 & 0 & 0 & 0 \\
0 & 0 & 0 & 0 & 1 & 0 \\
0 & 0 & 0 & 0 & 0 & 1 \\
\end{pmatrix}.
\end{equation}
It performs the following mapping: $p_x \rightarrow p_y$, $p_y
\rightarrow 1-p_x$ and $p_z \rightarrow p_z$. We conclude that
2-entropy is the only $\alpha$-entropy that is conserved under
continuous transformations. The transformations are orthogonal
matrices in the three-dimensional space of ``spin'' mean values
$(2p_x-1,2p_y-1,2p_z-1)$. We note that the requirement of
conservation of the measure of information under continuous
transformation is closely related to the continuity axiom in the
formulation of Hardy~\cite{hardy1}: ``There exists a continuous
reversible transformation on a system between any two pure states of
that system for systems of any dimension $N$.''

In Ref.~\cite{bruknerzeilinger} we showed that 2-entropy has the
property of invariance and in Ref.~\cite{shannoncritics} that
Shannon's measure does not have this property, but the general proof
of uniqueness was lacking.

\section{Higher-dimensional systems: any post-selected two-dimensional system contains at most one bit of information}

We now give an interesting observation on the scaling of the number
of degrees of freedom $K(N)$ with dimension $N$. We assume that it
is possible to perform a {\it selective} measurement on a
higher-dimensional system to prepare a genuine two-dimensional
system. Suppose that two-dimensional system is the most fundamental
one in the following particular sense:

(iv) ({\bf Higher-dimensional Systems}): Allowed physical states of
a general $N$ dimensional system are those and only those from which
one can generate allowed physical states of two dimensional systems
in post-selected measurements (i.e. systems that contain at most one
bit of information).

As mentioned above the parameter counting argument based on
considerations of composite systems requires that $K(N)=N^r-1$ with
$r \in \mathbf{N}$. Suppose that a single two-dimensional state
requires $K(2):=m$ independent measurements (e.g. mutually
complementary ones) to be described completely. One can now think
about all possible independent measurements that can be performed on
all possible two-dimensional systems that can be post-selected from
the full system of dimension $N$. From the selective measurement one
obtains $N-1$ independent probabilities (the sum of all of them is
unity). For each pair $(i,j)$ of outcomes in the selective
measurement one constructs $m-1$ additional measurements, to
describe the two-dimensional system completely. This gives
altogether
\begin{equation}
K(N)= N-1 + \frac{1}{2} N (N-1) (m-1),
\end{equation}
which is equal to $N^r-1$ only for $r=2$ and $m=3$. This is the case
of quantum theory.

\begin{figure}
  \includegraphics[width=0.70\textwidth]{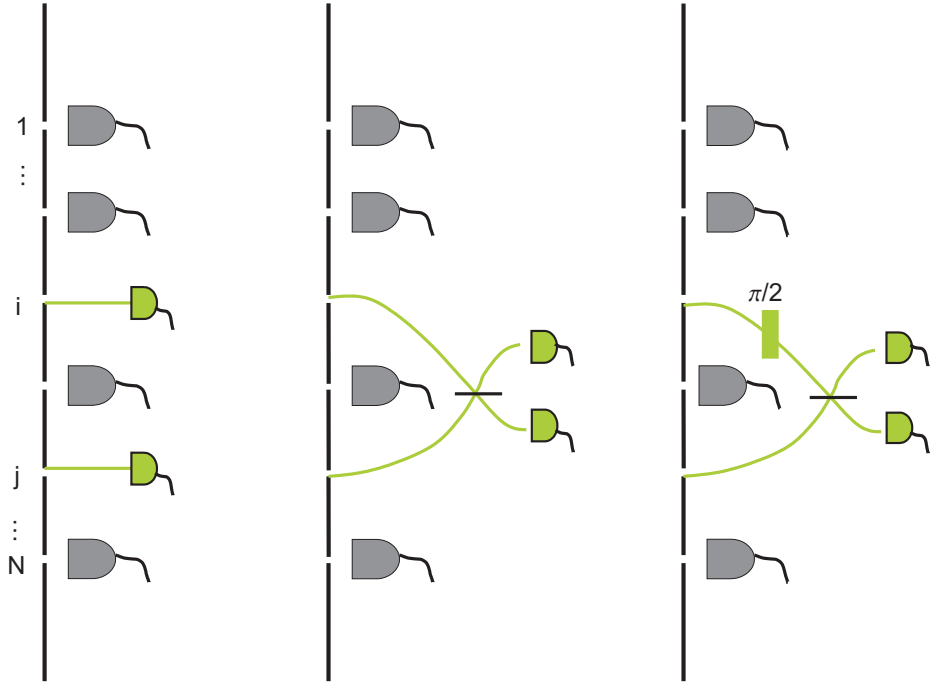}
\caption{Measurements to post-select a two-dimensional system. If
none of the grey detectors registers an event, a two-dimensional
system is prepared by post-selection from a higher-dimensional one.
After post-selection one can perform a complete set of three
mutually complementary measurements. In the figure the three
measurements are (1) which-path measurement; (2) interference
experiment with zero relative phase between two path states and (3)
interference experiment with $\pi/2$ relative phase between two path
states.}
\label{fig:2}       
\end{figure}

Taking scaling $K(N)=N^2-1$ as given we represent the state by a
probability vector whose first $N$ components are chosen to be
probabilities $p_1,...,p_N$ of $N$ distinguishable outcomes in a
single measurement. This measurement will be called $Z$ measurement.
In addition, for each pair $(i,j)$ of outcomes of the $Z$
measurement one constructs two additional measurements $X_{ij}$ and
$Y_{ij}$. These two measurements have all outcomes the same as the
$Z$ measurement except for the pair of outcomes $(i,j)$. From the
two additional measurements one obtains two additional independent
probabilities $p_{xij}$ and $p_{yij}$ for each pair $(i,j)$. This
gives altogether $N^2-1$ independent parameters as required. The
probability vector thus can be given as ${\bf p}=
(p_1,...,p_N,...,p_{xij},p_{yij},...)$. Next, consider setting up
the $Z$ measurement with $N$ distinguishable outcomes and placing
$N-2$ detectors to register all but the $i$ and $j$ outcomes as
shown in the figure 1. One then selects all those cases when no
detector registers an event. We assume that in this post-selected
manner one can prepare an arbitrary two-dimensional system
(elementary system). The two additional measurements described above
together with the measurement with outcomes $i$ and $j$ constitute a
complete set of mutually complementary measurements for the
two-dimensional system. In the subensemble of cases when the
postselection occurs the total information contained in a complete
set of mutually complementary measurements in the subselected
elementary system is at most one bit of information. We note that
this requirement is closely related to Hardy's subspace
axiom~\cite{hardy1}: ``... all systems of dimension $N$; or systems
of higher dimension but where the state is constrained to an
$N$-dimensional subspace, have the same properties.''

Since we have performed a ``degenerative  measurement'' in which we
have not distinguished between outcomes $i$ and $j$ it is natural to
assume that the state in the two-dimensional sector does not change
in the measurement except for normalization. Thus, one has ${\bf
p}=\frac{1}{p_i+p_j} (p_i,p_j,p_{xij}, 1-p_{xij}, p_{yij},
1-p_{yij})$.

It follows from (iv) that the principle of finiteness of information
(ii) -- the total information content of a two-dimensional system is
not larger than one bit of information -- should also apply to all
possible two-dimensional systems that can be post-selected from a
higher-dimensional system. We will show that this requirement is
equivalent to the positivity of the density operator in the Hilbert
space formulation of quantum theory. To prove this it is crucial to
use the quadratic measure of information in the expression for the
total uncertainty:
\begin{eqnarray}
H_{total}&=&
H_{2}\left(\frac{p_i}{p_i+p_j},\frac{p_j}{p_i+p_j}\right)+H_{2}\left(\frac{p_{xij}}{p_i+p_j},1-\frac{p_{xij}}{p_i+p_j}\right)+
H_{2}\left(\frac{p_{yij}}{p_i+p_j},1-\frac{p_{yij}}{p_i+p_j}\right)
\nonumber
\\ &=& \frac{4}{(p_i+p_j)^2} [ p_ip_j+ p_{xij}(1-p_{xij}) +
p_{yij}(1-p_{yij})] \geq 2. \label{tunel}
\end{eqnarray}
This is valid for any choice of measurements $Z$, $X_{ij}$ and
$Y_{ij}$. Here we put $k=2$ in the definition of 2-entropy to
guarantee the proper normalization as in Eq.~(\ref{invariance}).

We now show that a set of conditions~(\ref{tunel}) applied to all
$(i,j)$, $i \neq j \in \{1,...,N \}$, and for all possible complete
$Z$ measurements is equivalent to positivity of density operators in
quantum theory. We follow the idea of the proof of Ref.~\cite{nha}.
Consider a general Hermitian operator $\hat{\rho}$ of arbitrary
dimension $N$ with unit trace \footnote{The unit trace condition can
be relaxed if instead one requires a positivity of variance. See
Ref.~\cite{nha}.} Tr$\{\hat{\rho}\} = 1$. The operator has real
diagonal elements and due to unit trace condition there always
exists at least one positive diagonal element. Suppose that we have
performed the post-selected measurement as introduced above. Then
the density operator of the two-dimensional system is $\hat{\rho} =
\frac{1}{\rho_{ii} + \rho_{jj}}
\begin{pmatrix}
\rho_{ii}  & \rho_{ij} \\
\rho_{ji} &  \rho_{jj} \\
\end{pmatrix}$. We choose for the set of mutually complementary
measurement the three pseudo-spins: $\hat{\sigma}_z=|i\rangle
\langle i| -|j\rangle \langle j|$, $\hat{\sigma}_x=|i\rangle \langle
j| +|j\rangle \langle i|$ and $\hat{\sigma}_y=i (-|i\rangle \langle
j| + |j\rangle \langle  i|)$. Calculating the corresponding
probabilities in Eq.~(\ref{tunel}) one obtains $\rho_{ii} \rho_{jj}
\geq |\rho_{12}|^2 \geq 0$ for {\it all} choices of $i$ and $j$.
This with the condition of unit trace implies that {\it all}
diagonal elements are non-negative, i.e. $\rho_{ii} \geq 0$ for all
$i$. Since the condition on the total uncertainty applies to all
possible choices of the $Z$ measurement, this shows that the density
operator is positive.

Concluding this section we note that information invariance (iii)
together with the homogeneity of the parametric space leads to the
well known cosine-law (Malus' law) for quantum
probabilities~\cite{bruknerzeilinger}. Again it is crucial to use
quadratic measure of information.

\section{Conclusions}

We think that there are sufficient evidences which indicate that so
far hidden-variable approach could not encourage any new
phenomenology that might result in the hope for a progressive
research program towards answering Wheeler's famous question ``Why
the quantum?''. We suggest that a clear and promising alternative to
hidden-variable approach is to consider theories in which
probabilities have irreducible character. A guiding idea that we
follow here in order to impose a non-trivial structure on a
probabilistic theory is the principle that {\it a two-dimensional
system contains one bit of information.}

The assumption of invariance and continuity, i.e. that this one bit
of information is conserved under the continuous change of a set
mutually complementary measurements, uniquely singles out that
specific measure of information which is quadratic in probabilities.
This says, however, nothing about the scaling $K(N)$ of the number
of degrees of freedom with the number of dimension $N$ of a system.
If, however, one requires that {\it all} physical states of an $N$
dimensional system are those and only those from which one can
post-select physical two-dimensional states, one obtains the scaling
$K(N)=N^2-1$ as in quantum mechanics, taking in addition the
parameter-counting argument of Wootters~\cite{wootters} and of
Hardy~\cite{hardy} into account.

The principle of limited information is assumed to apply to every
two-dimensional systems and therefore also to all such post-selected
two-dimensional systems. Using the measure of information that is
quadratic in probabilities one can show that this requirement is
equivalent to the positivity of the density operator in quantum
mechanics.

\begin{acknowledgements}
We acknowledge discussions with B. Dakic and T. Paterek. This work
is supported by the Austrian Science Foundation within SFB (P08),
FWF Project No. P19570-N16 and Doctoral Program CoQuS, by the
Foundational Questions Institute (FQXi) and ERC Advanced
Investigator Grant (Project Number 227844).
\end{acknowledgements}

\end{document}